\documentclass[conference,10pt]{IEEEtran}
\IEEEoverridecommandlockouts
\usepackage{cite}
\usepackage{amsmath,amssymb,amsfonts}
\usepackage{algorithmic}
\usepackage{graphicx}
\usepackage{subfigure}
\usepackage{epstopdf}
\usepackage{textcomp}
\usepackage{xcolor}
\usepackage{mathtools}
\usepackage{array}
\def\BibTeX{{\rm B\kern-.05em{\sc i\kern-.025em b}\kern-.08em
    T\kern-.1667em\lower.7ex\hbox{E}\kern-.125emX}}
\begin{document}

\title{A Received Power Model for \\
Reconfigurable Intelligent Surface \\
and Measurement-based Validations
}

\author{\IEEEauthorblockN{Zipeng Wang\IEEEauthorrefmark{1}, Li Tan\IEEEauthorrefmark{1}, Haifan Yin\IEEEauthorrefmark{1}, Kai Wang\IEEEauthorrefmark{1}, Xilong Pei\IEEEauthorrefmark{1}, David Gesbert\IEEEauthorrefmark{2}}

\IEEEauthorblockA{\IEEEauthorrefmark{1}Huazhong University of Science and Technology, Wuhan, China}
\IEEEauthorblockA{\IEEEauthorrefmark{2}EURECOM, Sophia-Antipolis, France}
Email: {wangzipeng0421@gmail.com, \{ltan, yin, kaiw, pei\}@hust.edu.cn}, gesbert@eurecom.fr}

\maketitle

\begin{abstract}
The idea of using a Reconfigurable Intelligent Surface (RIS) consisting of a large array of passive scattering elements to assist wireless communication systems has recently attracted much attention from academia and industry. A central issue with RIS is how much power they can effectively convey to the target radio nodes. Regarding this question, several power level models exist in the literature but few have been validated through experiments. In this paper, we propose a  radar cross section-based received power model for an RIS-aided wireless communication system that is rooted in the physical properties of RIS. Our proposed model follows the intuition that the received power is related to the distances from the transmitter/receiver to the RIS, the angles in the TX-RIS-RX triangle, the effective area of each element, and the reflection coefficient of each element. To the best of our knowledge, this paper is the first to model the angle-dependent phase shift of the reflection coefficient, which is typically ignored in existing literature. We further measure the received power with our experimental platform in different scenarios to validate our model. The measurement results show that our model is appropriate both in near field and far field and can characterize the impact of angles well.
\end{abstract}

\begin{IEEEkeywords}
Reconfigurable intelligent surface, RIS, IRS, passive scattering elements, received power model, validations
\end{IEEEkeywords}

\section{Introduction}

The fifth-generation (5G) wireless networks are under deployment over the world. As one of the key technologies in 5G, massive multiple-input-multiple-output (mMIMO)  makes use of a large number of antennas (e.g., $\geq 64$) to increase the spectral efficiency (SE) by, at least, an order of magnitude compared to 4G \cite{marzetta:10a}. Despite this, both academia and industry continue their efforts to allow meeting ever more stringent requirements, such as ultra high data rate and energy efficiency, global coverage and connectivity, as well as extremely high reliability and low latency \cite{Saad2020Vision}. To solve these challenges, different avenues are considered, such as to further densify the network, using even ultra-massive MIMO arrays or utilizing the large and available bandwidth of higher frequency bands such as (sub-) terahertz frequencies \cite{Boccardi2014Five} \cite{Rajatheva2020White}. However, all these methods lead to higher energy consumption. Moreover, one of the major weaknesses for higher frequency bands is the limited network coverage, which is due to the high penetration loss and the lack of scattering and diffraction. 

In parallel, a novel concept named reconfigurable intelligent surface has received more and more attention from academia and industry \cite{Qingqing2020Towards}. It is also known as intelligent reflecting surfaces, large intelligent surfaces or software-controlled metasurfaces. Briefly speaking, RIS is a two-dimensional (i.e., with near-zero thickness) surface consisting of a large array of passive scattering elements, each of which can be controlled in a software-defined manner to impose an amplitude change and/or phase shift to the incident signals. By tuning the reflecting phases of the incident signals independently, the reflected signals can be added coherently to improve the received signal power or combined destructively to mitigate interference when arriving at the user equipment (UE) \cite{Shiming2020Toward}. In general, RIS is a prospective paradigm to realize a smart and reconfigurable wireless environment and improve the transmission performance with low energy consumption and hardware cost.

So far, numerous investigations have been performed on different aspects of RIS, such as prototyping \cite{Tang2019Wireless}\cite{Dai2020} \cite{Pei2021Ris-aided}, channel estimation  \cite{Zheng2020Intelligent}, promising application scenarios \cite{Liang2019Large}, etc. Among them, the modeling of received power for RIS-aided wireless communication systems is a fundamental yet challenging work, which is of great significance to the understanding of the RIS technology itself. Although several received power models have been proposed, most of them are lacking in experimental validations. The authors of \cite{Emil2020Power} make use of electromagnetic arguments to derive an exact channel gain expression that captures three essential near-field behaviors. 
\cite{Ozdogan2020Intelligent} takes advantage of the physical optics techniques to derive the far-field pathloss model that is inversely proportional to the square of the product of distances. In \cite{Ellingson2019Path}, the received signal power is formulated as the sum of each scattering element's received power which is related to the distances between the RIS elements and the transceivers, the element gains, and the complex-valued coefficients of the elements. The authors in \cite{Tang2021Wireless} develop a free-space pathloss model for RIS and experimental validations were also performed. Their proposed pathloss model is related to the distances from transceivers to the RIS, the size of the RIS, and the radiation patterns of antennas and elements. 

In this paper, we propose a novel radar cross section (RCS)-based received power model. 
To the best of our knowledge, our model is the first that takes angles of incidence/reflection into account when describing the phase shifts of the reflection elements. More specifically, the reflection phase in our model is not only a function of the control voltage imposed on the RIS element, but also a function of the angles. This is an important fact to consider when calculating the desired reflection coefficients. 
We validate the proposed model on our hardware platform and show that the simulation results based on our proposed model is very close to measurements. 

\section{System Model}
\begin{figure}[!htbp]
	\centering
	\includegraphics[width=1\linewidth]{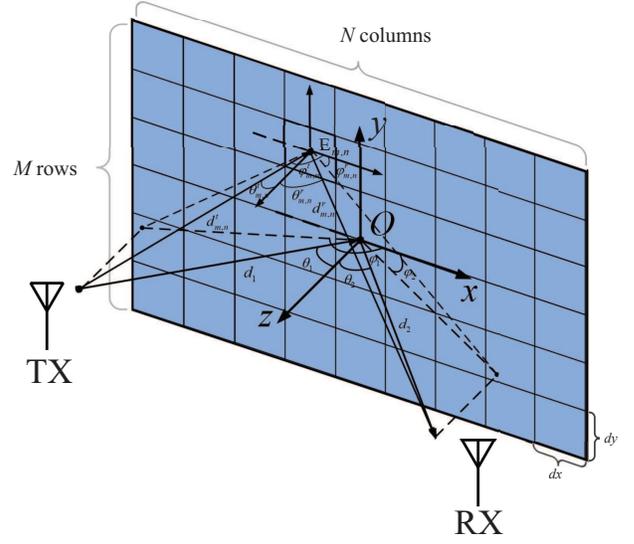}
	\caption{The geometric illustration of an RIS-aided communication system.}
	\label{Geometric}
\end{figure}

We consider an RIS-aided wireless communication system as shown in Fig.~\ref{Geometric}. For simplicity, the transmitter and receiver are both equipped with a single antenna. However, the results of this paper can be easily generalized to the case of MIMO. 

Since the model of line of sight (LOS) propagation  between the transmitter and the receiver is well known, we focus on modeling the signals reflected by the RIS in this paper only. In case the LOS path exists, we just need to add the signal of the LOS path and the reflected paths. 
Referring to Fig.~\ref{Geometric}, the RIS with $M \times N$ elements is placed on the $x$-$y$ plane of a Cartesian coordinate system and each row is parallel to the $x$-axis. The origin of the coordinate system is located at the geometric center of the RIS. Let $\text{E}_{m,n}$ denote the element in the $m$-th row and $n$-th column. The length and width of an element are denoted by $dx$ and $dy$, respectively. The location of $\text{E}_{m,n}$ is expressed as 
\begin{equation}
    \label{eq:pmn}
    \boldsymbol{p}_{m,n}=\left( \left( n-\frac{N+1}{2} \right)dx, \left( \frac{M+1}{2}-m \right)dy, 0 \right)
\end{equation}
in the Cartesian coordinate system.

Furthermore, we use the spherical coordinates to represent the positions of the transmitter/receiver. Specifically, $\left(d_{1},\theta_{1},\varphi_{1}\right)$ and $\left(d_{2},\theta_{2},\varphi_{2}\right)$ denote respectively the positions of the transmitter and receiver seen from the geometric center of the RIS plane, where $d,\theta,\varphi$ represent the distance, zenith angle and azimuth angle, respectively. If seen from $\text{E}_{m,n}$, the positions of the transmitter and receiver can be expressed as $\left(d_{m,n}^{t},\theta_{m,n}^{t},\varphi_{m,n}^{t}\right)$ and  $\left(d_{m,n}^{r},\theta_{m,n}^{r},\varphi_{m,n}^{r}\right)$. The spherical coordinates can be transformed into Cartesian coordinates by
\begin{equation}
	\label{eq:shptocar}
	\left\{
		\begin{array}{l}
			C_x = d \sin \theta \cos \varphi,\\
			C_y = d \sin \theta \sin \varphi,\\
			C_z = d \cos \theta.
		\end{array}
	\right.
\end{equation}

Denote the transmit signal as $x$. The received signal is written as
\begin{equation}
	\label{eq:received_signal} 
	y = \left(\sum_{m=1}^{M} \sum_{n=1}^{N} h_{m,n} g_{m,n}  \Gamma_{m,n}\right) x + z,
\end{equation}
where $h_{m,n}$ is the channel coefficient between the transmitter and the reflecting element $\text{E}_{m,n}$, and $g_{m,n}$ is the channel coefficient between $\text{E}_{m,n}$ and the receiver. $ z $ is the noise. $\Gamma_{m,n}$ is the reflection coefficient of the element $\text{E}_{m,n}$. 

\section{The Received Power Model}
In this section, we propose a general model for the received power in the RIS-aided communication system. 
Denote the transmit power by $P_{t}$ and transmit antenna gain by $G_{t}\left(\textbf{r}_{m,n}^{t}\right)$, where $\textbf{r}_{m,n}^{t}$ represents the direction from the transmitter to the element $\text{E}_{m,n}$ and is related to $\theta_{m,n}^{t}$ and $\varphi_{m,n}^{t}$. Then the spatial power density generated by the transmitter on the element $\text{E}_{m,n}$ is 
\begin{equation}
	\label{eq:Smnt}
	S_{m,n}^{t}=P_{t} G_{t}\left(\textbf{r}_{m,n}^{t}\right) \frac{1}{4 \pi \left( {d_{m,n}^{t}}\right) ^{2}},
\end{equation}
where $d_{m,n}^{t}$ represents the distance from the transmitter to $\text{E}_{m,n}$. With the transmitter located at $\left(C_{x},C_{y},C_{z}\right)$, $d_{m,n}^{t}$ can be calculated by 
\begin{equation}
	\label{eq:rmnt}
	\resizebox{.86\hsize}{!}{$d_{m,n}^{t}=\sqrt{\left[ C_{x}\!-\! \left( n\!-\!\frac{N\!+\!1}{2} \right)\!dx \right] ^{2} + \left[ C_{y}\!-\! \left( \frac{\!M\!+1}{2}\!-\!m \right)\!dy \right] ^{2} +C_{z}^{2}}.$}
\end{equation}

The power captured by $\text{E}_{m,n}$ is 
\begin{equation}
	\label{eq:Pmnt}
	P_{m,n}^{t}=S_{m,n}^{t}A_{e}\left(-\textbf{r}_{m,n}^{t}\right),
\end{equation}
where $A_{e}\left(-\textbf{r}_{m,n}^{t}\right)$ is the effective area of the $m,n$-th element in the direction $-\textbf{r}_{m,n}^{t}$. Similar to \eqref{eq:Smnt}, the spatial power density generated by the $m,n$-th element is
\begin{equation}
	\label{eq:Smnr}
	S_{m,n}^{r}=P_{m,n}^{t} G_{e}\left(\textbf{r}_{m,n}^{r}\right) \frac{1}{4 \pi \left( {d_{m,n}^{r}}\right) ^{2}},
\end{equation}
where $d_{m,n}^{r}$ represents the distance from $\text{E}_{m,n}$ to the receiver, which can be calculated similarly to \eqref{eq:rmnt}. $G_{e}\left(\textbf{r}_{m,n}^{r}\right)$ is the gain of the element $\text{E}_{m,n}$ in the direction $\textbf{r}_{m,n}^{r}$. The relation between the effective area and element gain is generally expressed as \cite{Gustafsson2019Maximum}
\begin{equation}
	\label{eq:G_A}
	G_{e}\left( \textbf{r}_{m,n}^{r} \right) = \frac{4 \pi }{\lambda^2} A_{e}\left( \textbf{r}_{m,n}^{r}\right).
\end{equation}
Note that we do not consider the power loss of the reflection in \eqref{eq:G_A}. However, the efficiency of the reflection will later be taken into account in the reflection coefficient. 
The power captured by the receiver that is reflected by the $m,n$-th element is 
\begin{equation}
	\label{eq:Pmnr}
	P_{m,n}^{r}=S_{m,n}^{r}A_{r}\left(-\textbf{r}_{m,n}^{r}\right),
\end{equation}
where $A_{r}\left(-\textbf{r}_{m,n}^{r}\right)$ is the effective area of the receiving antenna in the direction $-\textbf{r}_{m,n}^{r}$, which has the similar relation with antenna gain $G_{r}\left(-\textbf{r}_{m,n}^{r}\right)$, namely
\begin{equation}
	\label{eq:A_G}
	A_{r}\left( -\textbf{r}_{m,n}^{r} \right) = \frac{\lambda^2}{4 \pi \eta_{r}} G_{r}\left(-\textbf{r}_{m,n}^{r} \right),
\end{equation}
where $\eta_{r} \in \left[ 0,1\right] $ is the efficiency of the receiving antenna. According to \eqref{eq:Smnt} $\thicksim$ \eqref{eq:A_G}, the received power reflected from the $m,n$-th element is 
\begin{equation}
	P_{m,n}^{r}=P_{t} G_{t}\left(\textbf{r}_{m,n}^{t}\right)G_{r}\left(-\textbf{r}_{m,n}^{r} \right) \frac{A_{e}\left(-\textbf{r}_{m,n}^{t}\right) A_{e}\left( \textbf{r}_{m,n}^{r}\right)}{\left( 4 \pi d_{m,n}^{t} d_{m,n}^{r} \right)^{2}\eta_{r}}.
\end{equation}

Some recent experiments in \cite{Pei2021Ris-aided} have indicated that both the amplitude and phase shift of the reflection coefficient are related to the angle of incidence, the angle of reflection, and the control signal, e.g., the bias voltage of a varactor. An example is that, for a given control signal, the reflection coefficient also varies with the angles of arrival and departure.  As a result, we model the reflection coefficient as a function of the three factors mentioned above. For the element $\text{E}_{m,n}$, the reflection coefficient is modeled as
\begin{align}
	\label{eq:Gamma_mn}
	& \Gamma_{m,n}\left( -\textbf{r}_{m,n}^{t}, \textbf{r}_{m,n}^{r}, u_{m,n}\right)  \\
	& = \mu_{m,n}\left( -\textbf{r}_{m,n}^{t}, \textbf{r}_{m,n}^{r}, u_{m,n}\right) \mathrm{e}^{\text{j} \phi_{m,n}\left( -\textbf{r}_{m,n}^{t}, \textbf{r}_{m,n}^{r}, u_{m,n}\right)}, \nonumber
\end{align}
where $\mu_{m,n}\left( -\textbf{r}_{m,n}^{t}, \textbf{r}_{m,n}^{r}, u_{m,n}\right)$, $\phi_{m,n}\left( -\textbf{r}_{m,n}^{t}, \textbf{r}_{m,n}^{r}, u_{m,n}\right)$ represent the amplitude  attenuation and phase shift, respectively. $u_{m,n}$ is the control signal imposed on the element $\text{E}_{m,n}$. 

The total received power reflected by the RIS is then written as
\begin{equation}
	\label{eq:Pr}
	P_{r} = \left| \sum_{m=1}^{M} \sum_{n=1}^{N} \Gamma_{m,n}\left( -\textbf{r}_{m,n}^{t}, \textbf{r}_{m,n}^{r}, u_{m,n}\right)  \sqrt{P_{m,n}^{r}} \mathrm{e}^{\text{j} \Phi_{m,n}} \right| ^{2},
\end{equation}
where $\Phi_{m,n}$ is the phase difference of the $m,n$-th element caused by the propagation delay:
\begin{equation}
    \Phi_{m,n} =\frac{2\pi}{\lambda}\left(d_{m,n}^{t}+d_{m,n}^{r} \right).
\end{equation}

In order to make the expression more intuitive, the concept of radar cross section (RCS) $\sigma_{m,n}\left( -\textbf{r}_{m,n}^{t}, \textbf{r}_{m,n}^{r}, u_{m,n}\right)$ is introduced to describe the joint impact of the amplitude attenuation and the effective area of the element $\text{E}_{m,n}$, namely
\begin{align}
		\label{eq:sigma_mn}
		&\sigma_{m,n}\left( -\textbf{r}_{m,n}^{t}, \textbf{r}_{m,n}^{r}, u_{m,n}\right)  \\ 
		& = \mu_{m,n}\left( -\textbf{r}_{m,n}^{t}, \textbf{r}_{m,n}^{r}, u_{m,n}\right) \sqrt{A_{e}\left(-\textbf{r}_{m,n}^{t}\right) A_{e}\left( \textbf{r}_{m,n}^{r}\right)}. \nonumber
\end{align}
The proposed received power model can now be written as
\begin{equation}
    \label{eq:Pr_appr}
	P_{r} =  \frac{P_{t}}{16 \pi^{2} \eta_{r}} \left| \sum_{m=1}^{M} \sum_{n=1}^{N} \frac{\sqrt{G_{t} G_{r}} \sigma_{m,n} \mathrm{e}^{\text{j} \left(\phi_{m,n} + \Phi_{m,n}\right)}}{d_{m,n}^{t}d_{m,n}^{r}} \right| ^{2}. 
\end{equation}
Note that we temporarily drop the arguments of $G_{t}$, $G_{r}$, $\sigma_{m,n}$ and $\Phi_{m,n}$ in order to make the above expression more concise.

We now study the impact of angles on the RCS, which depends on multiple parameters, e.g., the incident/reflection angles and the efficiency of reflection. In practice, such a multi-dimensional data can be obtained by electromagnetic simulate software. 
For simplicity however, we let the angles of $\theta^{t}$ and $\varphi^{t}$ both equal to zero degree(also the case in our experiments), and characterize the impact of zenith angle of reflection $\theta^{r}$ on the RCS $\hat{\sigma}$.
we adopt the model of \cite{Meyers2003Encyclopedia} here: 
\begin{equation}
    \label{eq:RCS_fit}
	\hat{\sigma}\left(\theta^{r}\right) = \frac{4\pi A^{2}}{\lambda^{2}} \left[ \frac{\sin{\left( k\sqrt{A}\sin{(\theta^{r})}\right)}}{k\sqrt{A}\sin{(\theta^{r})}} \right]^{2} + c,
\end{equation}
where $A = dxdy$ is the geometry area of each element and $k=\frac{2\pi}{\lambda}$ is the wave number. $c$ is a constant which is related to other parameters, e.g., the reflection efficiency and/or the angle $\varphi^{r} $. 

We now characterize the impact of  $\theta^{r}$ on the phase shift. We propose to model such a relationship with a cosine function: 
\begin{equation}
    \label{eq:phi_fit}
	\hat{\phi}\left(\theta^{r}\right) = a \cos{(\theta^{r})} + b,
\end{equation}
where $a,b$ are the design parameters. An example of their values is given below. 

Fig.~\ref{fit_curve} illustrates the reflection coefficient as a function of $\theta^{r}$ with azimuth angle $\varphi^{r}$ fixed to $45^{\circ}$. We can observe that our models fit the CST simulations well.
\begin{figure}[!htbp]
    \centering
    \subfigure[]{
	\label{RCS_fit}
d	\includegraphics[width=0.45\linewidth]	{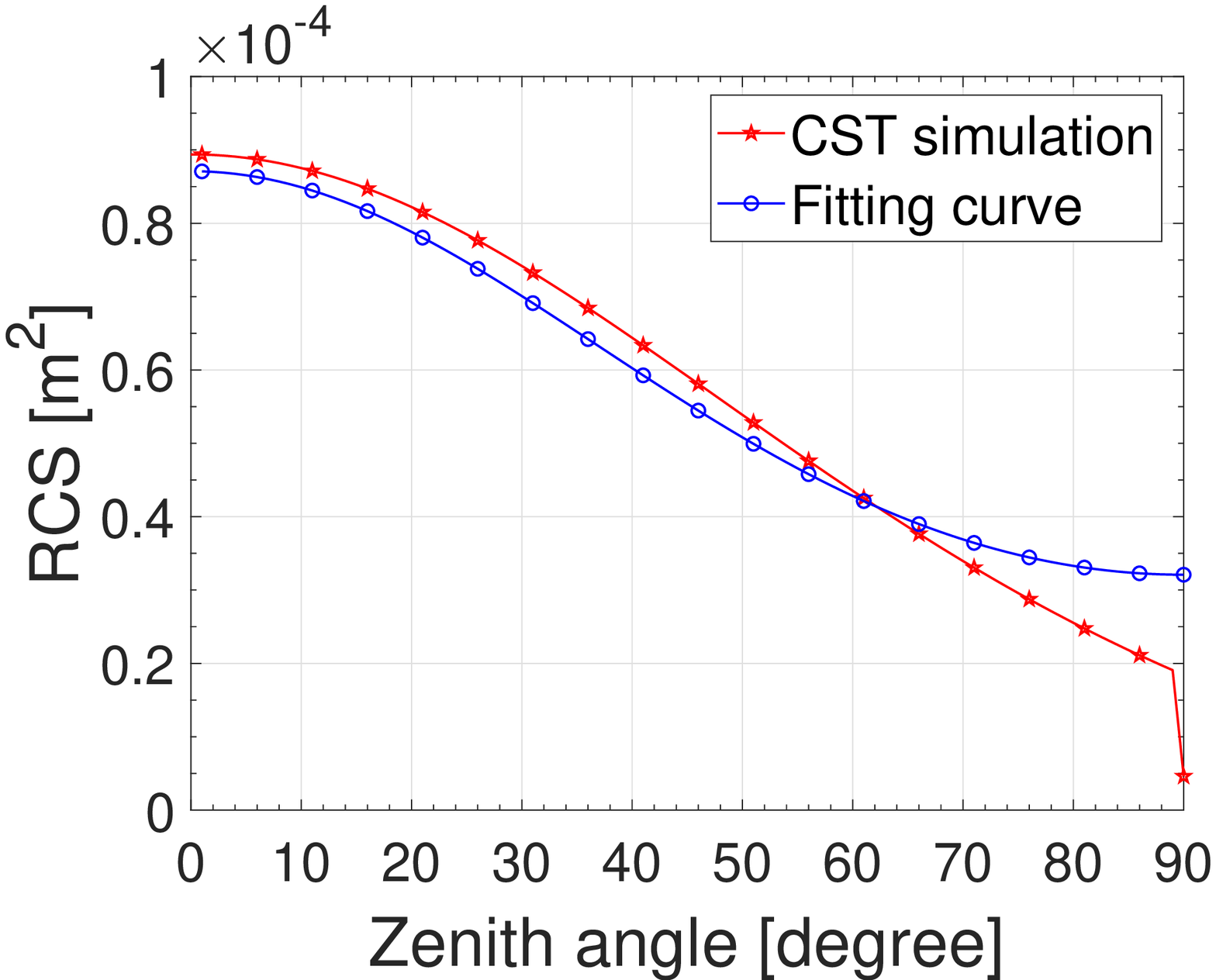}}
	\subfigure[]{
		\label{phase_fit}
		\includegraphics[width=0.45\linewidth] {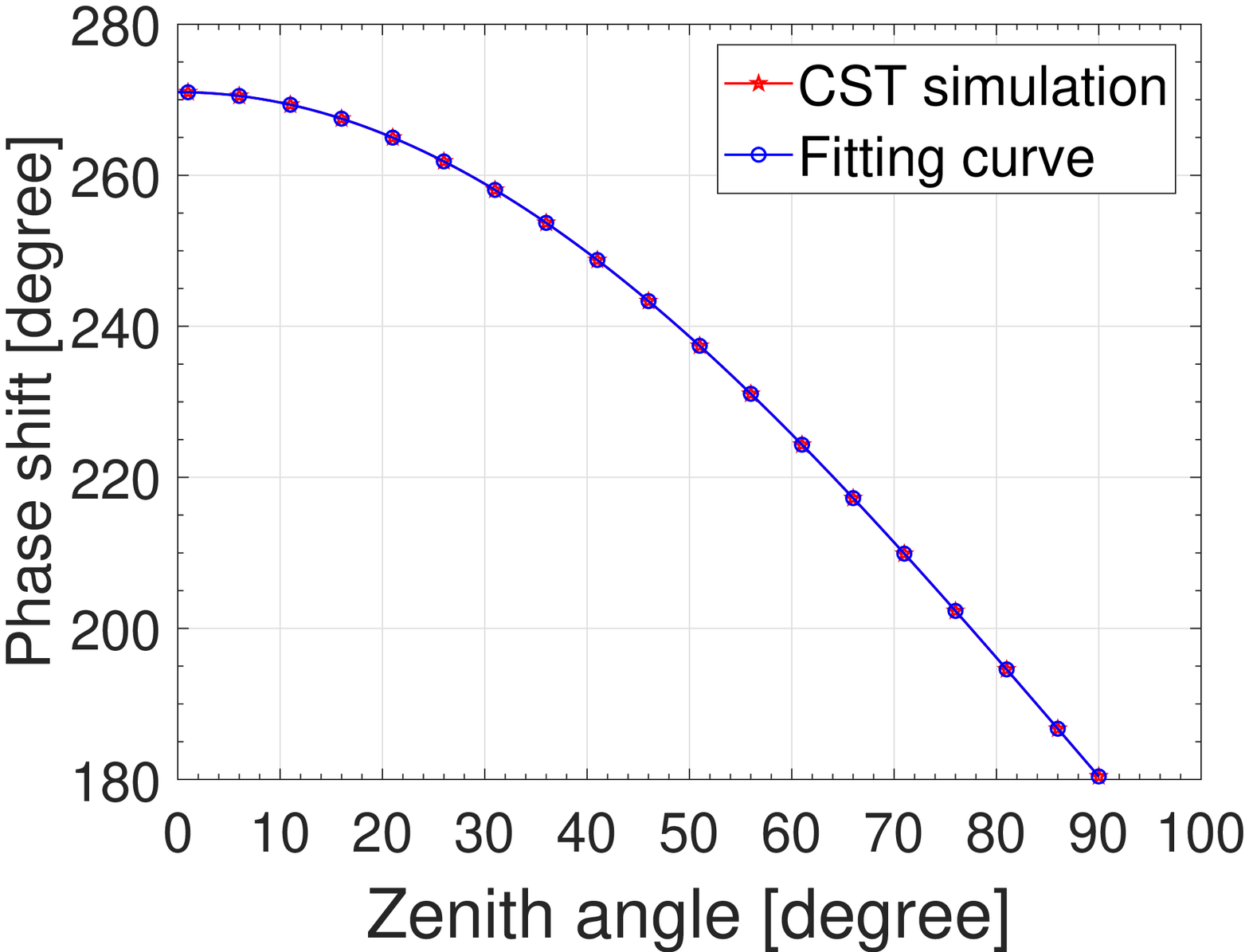}}
    \caption{The reflection coefficient versus the zenith angle $\theta^{r}$ with $\phi = 45^{\circ}$. (a) $\hat{\sigma}$ versus $\theta^{r}$ with $c\!=\!1.42\!\times\!10^{-5}~\text{m}^{2}$. (b) $\hat{\phi}$ versus $\theta^{r}$ with $a\!=\!90^{\circ},b\!=\!180^{\circ}$. }
    \label{fit_curve}
\end{figure}

\section{Measurements and Validations}
In this section, we validate the proposed model with measurements and compare it with an existing model. 
Some recent works have modeled the role of RIS as the specular reflection approximately when in the near field or the RIS is infinitely large \cite{Basar2019Wireless,Tang2021Wireless}. It means that the RIS acts as a plane mirror that reflects the incident signal with the same amplitude attenuation coefficient and perfect phase shift. Similar to the Friis transmission equation, the received power of the specular reflection model is expressed as 
\begin{equation}
	\label{eq:nearfield_Pr}
	P_{r} = P_{t} G_{t} G_{r} \left[ \frac{\lambda \bar{\mu}}{4 \pi \left( d_{1} + d_{2} \right)} \right]^{2},
\end{equation}
where $\bar{\mu}$ is the average amplitude of the reflection coefficient 
\begin{equation}
	\label{eq:barmu}
	\bar{\mu} = \frac{1}{MN}\sum_{m=1}^{M} \sum_{n=1}^{N} \mu_{m,n}\left( -\textbf{r}_{m,n}^{t}, \textbf{r}_{m,n}^{r}, u_{m,n}\right),
\end{equation}
and $G_{t}, G_{r}$ are the antenna gains. 

From \eqref{eq:nearfield_Pr}, one may observe that the received signal power is inversely proportional to $(d_{1} + d_{2})^2$. In the following we will validate this model and our proposed model \eqref{eq:Pr_appr}. 

The basic parameters of our experiments are listed in Table~\ref{basic}. Other details of our experimental platform can be found in \cite{Pei2021Ris-aided}.
\begin{table}[!htbp]
	\renewcommand\arraystretch{1.2}
	\caption{Basic Parameters Of the experiments}
	\begin{center}
		\begin{tabular}{c|c}
			\hline
			\textbf{Parameters}& \textbf{Value} \\
			\hline
			Operating frequency & $f = 5.8$ GHz  \\
			Antenna peak gain & $G_{t} = G_{r} = 17.1$ dB  \\
			Efficiency of antenna & $\eta_{r}  \approx 54.29 \%$  \\
			Rows and columns & $M = 20, N=55$  \\
			Element size & $dx = 14.3$ mm, $dy = 10.27$ mm  \\
			\hline
		\end{tabular}
		\label{basic}
	\end{center}
\end{table}

\begin{figure}[!htbp]
	\centering
	\subfigure[]{
	\label{nearfield}
	\includegraphics[width=0.95\linewidth]	{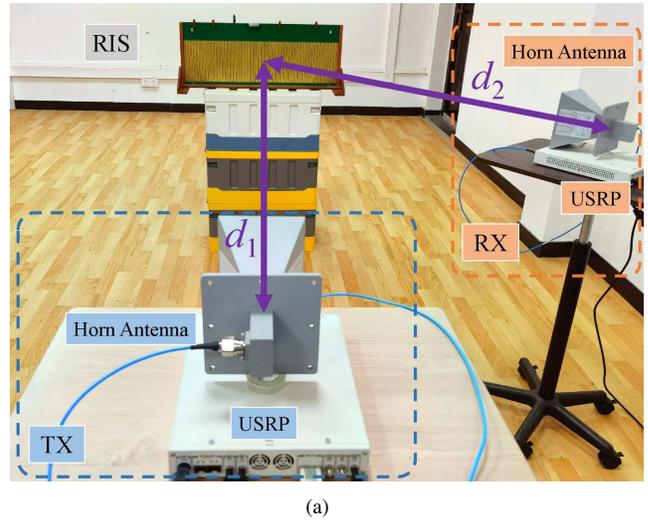}}
	\subfigure[]{
		\label{farfield}
		\includegraphics[width=0.95\linewidth] {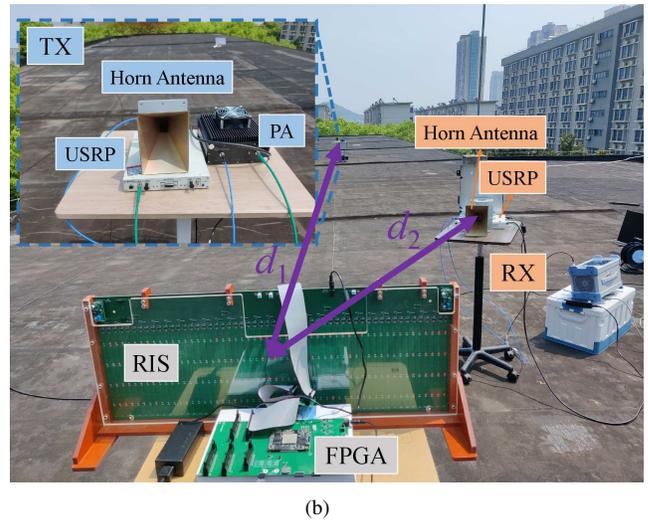}}
	\caption{Measurement environments. (a) Indoor near field testing scene. \\
	(b) Outdoor far field testing scene.}
	\label{Environments}
\end{figure}

Fig.~\ref{Environments} illustrates the indoor and outdoor measurement environments. 
The reflection coefficients are 1-bit quantized, i.e., the two control states have a phase difference of around 180 degrees, which is enabled by adjusting the bias voltage $u_{m,n}$ imposed to the varactor diodes. The reflection coefficients are adjusted according to Algorithm 1 in \cite{Pei2021Ris-aided}. 

We carry on near-field experiments indoor and far-field experiments outdoor. The boundary of near field and far field is defined as \cite{Tang2021Wireless}
\begin{equation}
	\label{eq:df}
	df = \frac{2MNdxdy}{\lambda^{2}}.
\end{equation}

For the RIS in our tests, the boundary $df \approx 6~\text{m}$. Thus, we select the distances from $1~\text{m}$ to $5~\text{m}$ in indoor near-field experiments and from $5~\text{m}$ to $50~\text{m}$ in outdoor far-field experiments. The antennas are placed in a way that the azimuth angles of the transmit/receive antennas seen from the origin of the RIS are always equal to zero degree(i.e., $\varphi_{1}=\varphi_{2}=0$).
\begin{figure}[!htbp]
	\centering
	\includegraphics[width=1\linewidth]{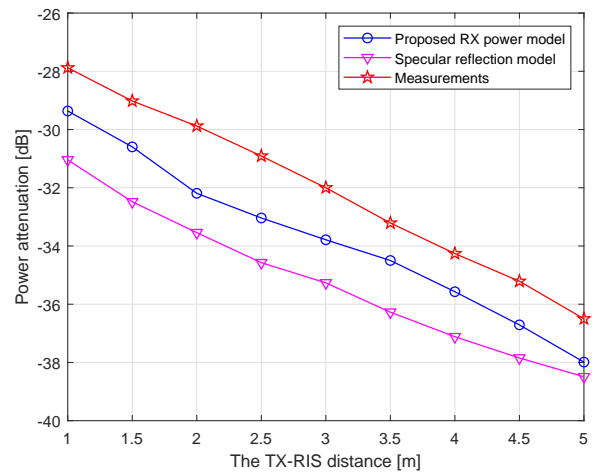}
	\caption{Power attenuation versus the TX-RIS distance $d_{1}$ with $d_{2}=2, \theta_{1}=0, \theta_{2}=30^{\circ}, \varphi_{1}=0, \varphi_{2}=0$. }
	\label{nearfield_distance}
\end{figure}

Fig.~\ref{nearfield_distance} illustrates the indoor near-field measurement results and the model results of the power attenuation $P_{r}/P_{t}$ as a function of the distance between the transmitter and the geometric center of the RIS (i.e., $d_{1}$). 
We can observe that the measurements have the same trend with our proposed RX power model, however with a gap of about $2$ dB. This difference might be due to the reflection of indoor metal objects and walls that makes the measurement results higher.
When it comes to the curve of the specular reflection model, it is also close to the measurement results in such a near-field case. These measurements validate the applicability of the received power model we proposed and the specular reflection model in this case.

\begin{figure}[!htbp]
	\centering
	\includegraphics[width=1\linewidth]{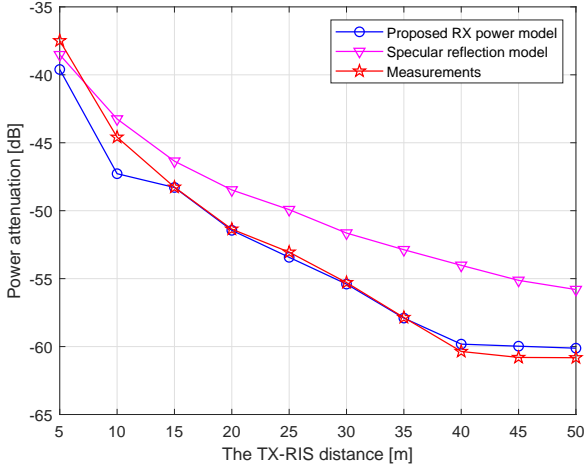}
	\caption{Power attenuation versus the TX-RIS distance $d_{1}$ with $d_{2}=2, \theta_{1}=0, \theta_{2}=30^{\circ}, \varphi_{1}=0, \varphi_{2}=0$. }
	\label{farfield_distance}
\end{figure}

Fig.~\ref{farfield_distance} illustrates the outdoor far-field results of the power attenuation. The horizontal axis represents the distance between the transmitter and the geometric center of the RIS (i.e., $d_{1}$).
We can observe that there is an increasing gap between the specular reflection model and the measurement results. However, simulation results based on our model are consistent with the measurements, which validates the accuracy of the proposed received power model in far-field case. 

\begin{figure}[!htbp]
	\centering
	\includegraphics[width=1\linewidth]{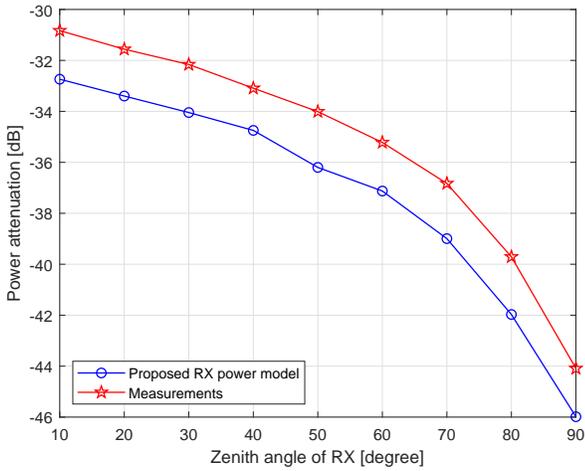}
	\caption{Power attenuation versus the zenith angle $\theta_{2}$ of RX with $d_{1}=3, d_{2}=2,\theta_{1}=0, \varphi_{1}=0, \varphi_{2}=0$. }
	\label{nearfield_angle}
\end{figure}

Fig.~\ref{nearfield_angle} illustrates the indoor power attenuation results as a function of the zenith angle of the receiver (i.e., $\theta_{2}$). We observe that the RX received power model we proposed matches the measurements with a gap of about $2$ dB. Since the specular reflection model does not consider the influence of angles, the received power model is a constant value, which is not aligned with the measured results. 
\section{Conclusion}
In this work, we have proposed a novel RCS-based received power model for an RIS-aided wireless communication system. This model introduces an angular-dependent phase shift behavior of the RIS elements, which has not yet been considered in previous models. We then validated it through indoor and outdoor measurements. 
The results show that the specular reflection model is able to describe the received power in near-field case when the incident/reflection angles are relatively small. However it fails when in the far-field case or in case the angles are large. By contrast, our model can hold for both near-field and far-field cases and it can characterize the impact of the angles well. 

\section*{Acknowledgment}
This work was supported by the National Natural Science 
Foundation of China under Grant 62071191. The work of D. Gesbert was partially funded via the HUAWEI France supported Chair on Future Wireless Networks at EURECOM.

\bibliographystyle{IEEEtran}
\bibliography{ref}

\begin{thebibliography}{10}
\providecommand{\url}[1]{#1}
\csname url@samestyle\endcsname
\providecommand{\newblock}{\relax}
\providecommand{\bibinfo}[2]{#2}
\providecommand{\BIBentrySTDinterwordspacing}{\spaceskip=0pt\relax}
\providecommand{\BIBentryALTinterwordstretchfactor}{4}
\providecommand{\BIBentryALTinterwordspacing}{\spaceskip=\fontdimen2\font plus
\BIBentryALTinterwordstretchfactor\fontdimen3\font minus
  \fontdimen4\font\relax}
\providecommand{\BIBforeignlanguage}[2]{{%
\expandafter\ifx\csname l@#1\endcsname\relax
\typeout{** WARNING: IEEEtran.bst: No hyphenation pattern has been}%
\typeout{** loaded for the language `#1'. Using the pattern for}%
\typeout{** the default language instead.}%
\else
\language=\csname l@#1\endcsname
\fi
#2}}
\providecommand{\BIBdecl}{\relax}
\BIBdecl

\bibitem{marzetta:10a}
T.~L. Marzetta, ``Noncooperative cellular wireless with unlimited numbers of
  base station antennas,'' \emph{IEEE Trans. Wireless Commun.}, vol.~9, no.~11,
  pp. 3590--3600, Nov. 2010.

\bibitem{Saad2020Vision}
W.~Saad, M.~Bennis, and M.~Chen, ``A vision of {6G} wireless systems:
  Applications, trends, technologies, and open research problems,'' \emph{IEEE
  Network}, vol.~34, no.~3, pp. 134--142, 2020.

\bibitem{Boccardi2014Five}
F.~{Boccardi}, R.~W. {Heath}, A.~{Lozano}, T.~L. {Marzetta}, and P.~{Popovski},
  ``Five disruptive technology directions for {5G},'' \emph{IEEE Commun. Mag.},
  vol.~52, no.~2, pp. 74--80, 2014.

\bibitem{Rajatheva2020White}
N.~Rajatheva, I.~Atzeni, E.~Bjornson, A.~Bourdoux, and W.~Xu, ``{White Paper on
  Broadband Connectivity in {6G}},'' \emph{arXiv e-prints}, p.
  arXiv:2004.14247, Apr. 2020.

\bibitem{Qingqing2020Towards}
Q.~{Wu} and R.~{Zhang}, ``Towards smart and reconfigurable environment:
  Intelligent reflecting surface aided wireless network,'' \emph{IEEE Commun.
  Mag.}, vol.~58, no.~1, pp. 106--112, 2020.

\bibitem{Shiming2020Toward}
S.~{Gong}, X.~{Lu}, D.~T. {Hoang}, D.~{Niyato}, L.~{Shu}, D.~I. {Kim}, and
  Y.~C. {Liang}, ``Towards smart wireless communications via intelligent
  reflecting surfaces: A contemporary survey,'' \emph{IEEE Commun. Surv.
  Tutor.}, vol.~22, no.~4, pp. 2283--2314, Fourthquarter 2020.

\bibitem{Tang2019Wireless}
W.~Tang, X.~Li, J.~Y. Dai, S.~Jin, Y.~Zeng, Q.~Cheng, and T.~J. Cui, ``Wireless
  communications with programmable metasurface: Transceiver design and
  experimental results,'' \emph{China Commun.}, 2019.

\bibitem{Dai2020}
L.~Dai, B.~Wang, M.~Wang, X.~Yang, J.~Tan, S.~Bi, S.~Xu, F.~Yang, Z.~Chen,
  M.~D. Renzo, C.-B. Chae, and L.~Hanzo, ``Reconfigurable intelligent
  surface-based wireless communications: Antenna design, prototyping, and
  experimental results,'' \emph{IEEE Access}, vol.~8, pp. 45\,913--45\,923,
  2020.

\bibitem{Pei2021Ris-aided}
X.~{Pei}, H.~{Yin}, L.~{Tan}, L.~{Cao}, Z.~{Li}, K.~{Wang}, K.~{Zhang}, and
  E.~{Bj{\"o}rnson}, ``{RIS-Aided Wireless Communications: Prototyping,
  Adaptive Beamforming, and Indoor/Outdoor Field Trials},'' \emph{arXiv
  e-prints}, p. arXiv:2103.00534, Feb. 2021.

\bibitem{Zheng2020Intelligent}
B.~Zheng and R.~Zhang, ``Intelligent reflecting surface-enhanced {OFDM}:
  Channel estimation and reflection optimization,'' \emph{IEEE Wireless Commun.
  Lett.}, vol.~9, no.~4, pp. 518--522, 2020.

\bibitem{Liang2019Large}
Y.-C. Liang, R.~Long, Q.~Zhang, J.~Chen, H.~V. Cheng, and H.~Guo, ``Large
  intelligent surface/antennas ({LISA}): Making reflective radios smart,''
  \emph{J. Commun. Netw.}, vol.~4, no.~2, pp. 40--50, 2019.

\bibitem{Emil2020Power}
E.~{Björnson} and L.~{Sanguinetti}, ``Power scaling laws and near-field
  behaviors of massive {MIMO} and intelligent reflecting surfaces,'' \emph{IEEE
  Open J. Commu. Soc.}, vol.~1, pp. 1306--1324, 2020.

\bibitem{Ozdogan2020Intelligent}
Özdogan Özgecan, E.~Björnson, and E.~G. Larsson, ``Intelligent reflecting
  surfaces: Physics, propagation, and pathloss modeling,'' \emph{IEEE Wireless
  Commun. Lett.}, vol.~9, no.~5, pp. 581--585, 2020.

\bibitem{Ellingson2019Path}
S.~W. {Ellingson}, ``{Path Loss in Reconfigurable Intelligent Surface-Enabled
  Channels},'' \emph{arXiv e-prints}, p. arXiv:1912.06759, Dec. 2019.

\bibitem{Tang2021Wireless}
W.~{Tang}, M.~Z. {Chen}, X.~{Chen}, J.~Y. {Dai}, Y.~{Han}, M.~{Di Renzo},
  Y.~{Zeng}, S.~{Jin}, Q.~{Cheng}, and T.~J. {Cui}, ``Wireless communications
  with reconfigurable intelligent surface: Path loss modeling and experimental
  measurement,'' \emph{IEEE Trans. Wireless Commun.}, vol.~20, no.~1, pp.
  421--439, Jan. 2021.

\bibitem{Gustafsson2019Maximum}
M.~Gustafsson and M.~Capek, ``Maximum gain, effective area, and directivity,''
  \emph{IEEE Trans Antennas Propag.}, vol.~67, no.~8, pp. 5282--5293, 2019.

\bibitem{Meyers2003Encyclopedia}
N.~Levanon, ``Radar,'' in \emph{Encyclopedia of Physical Science and Technology
  (Third Edition)}, 3rd~ed., R.~A. Meyers, Ed.\hskip 1em plus 0.5em minus
  0.4em\relax New York: Academic Press, 2003, pp. 497--510.

\bibitem{Basar2019Wireless}
E.~Basar, M.~Di~Renzo, J.~De~Rosny, M.~Debbah, M.-S. Alouini, and R.~Zhang,
  ``Wireless communications through reconfigurable intelligent surfaces,''
  \emph{IEEE Access}, vol.~7, pp. 116\,753--116\,773, 2019.

\end{thebibliography}
\end{document}